\pdfoutput=1
\documentclass[aps,twoside,twocolumn,pra,superscriptaddress]{revtex4-2}
\usepackage{epsfig,amsmath,amssymb}
\usepackage{graphicx}
\usepackage{array}
\usepackage{xcolor}
\usepackage[colorlinks=true,linkcolor=blue]{hyperref}

\def\ket#1{|#1\rangle}
\def\bra#1{\langle#1|}
\def\scal#1#2{\langle#1|#2\rangle}
\def\matr#1#2#3{\langle#1|#2|#3\rangle}

\def\={\!=\!}
\def\>{\!>\!}
\def\<{\!<\!}
\def\-{\!-\!}
\def\+{\!+\!}

\begin{document}

\title{Creating multicomponent Schr{\"o}dinger cat states in a coupled qubit-oscillator system}

\author{Pavel Str\'ansk\'y}
\email{pavel.stransky@matfyz.cuni.cz}
\affiliation{Institute of Particle and Nuclear Physics, Faculty of Mathematics and Physics, Charles University, V Hole\v{s}ovi\v{c}k\'ach 2, 18000 Prague, Czechia}
\author{Pavel Cejnar}
\email{pavel.cejnar@matfyz.cuni.cz}
\affiliation{Institute of Particle and Nuclear Physics, Faculty of Mathematics and Physics, Charles University, V Hole\v{s}ovi\v{c}k\'ach 2, 18000 Prague, Czechia}

\date{\today}

\begin{abstract}
We present a method for preparing various exotic modifications of Schr{\"o}dinger cat states by coupling a semiclassical oscillator to a system of qubits.
Varying the number of qubits and parameters of the protocol (involving quantum quench of the coupled system and a subsequent spin measurement), we bring the oscillator into a coherent superposition composed of an arbitrary number of wavepackets in tunable proportions and motion relations.
The method can be implemented with the aid of current experimental techniques and may find applications in quantum information and sensing protocols.
\end{abstract}

\maketitle

While in the early years of quantum theory the celebrated puzzle of Schr{\"o}dinger's cat \cite{Schr35}, which was thought in the~${\ket{\mathrm{dead}}\!+\!\ket{\mathrm{alive}}}$ quantum superposition state, served to metaphorically demonstrate the inherent strangeness of quantum laws, at present the so-called Schr{\"o}dinger-cat states \cite{Wine13,John17,Frow18,Omra19,Girv19,Hack19,Wang22,He23,Bild23,Bati24,Gupt24,Hosh25} play the role of an essential quantum resource.
Such states, whose laboratory realizations were reported on the platforms ranging from quantum optics to superconducting circuits and trapped ions, can be applied in quantum information and communication as robust realizations of qubits \cite{Niel00,Haro03,Mirr14,Fluh19,Grim20,Grav23,Yu25}, but also in quantum metrology due to the sub-Planck phase-space structures they typically exhibit \cite{Tosc06,Kwon19,Tats19}.

The original motivation for studying the cat-like states was to explore possibilities of creating {\em almost pure\/} superposition states in large (nearly macroscopic) quantum systems (see, e.g., Ref.\,\cite{Bild23}).
However, also the cat-like superpositions that show {\em maximal entanglement\/} of a large system with some ancillary qubits (so-called Bell cat states, in which the large system alone is in a fully incoherent mixed state) are of great importance from both fundamental and application points of view \cite{Vlas15,Lajc24}.
In this Letter, we present an implementable method for creating exotic cat-like states, which are somewhere in between the above two extremal cases.
Bringing the semiclassical system, namely an oscillator, into a {\em partially entangled\/} superposition state with a set of qubits, we generate highly non-classical behavior characterized by ${n\geq 2}$ macrosopically distinguishable and partly coherent wavepackets moving in the oscillator phase space (cf.\,Refs.\,\cite{Jans93,Lee94,Shuk19,Howa19,Hail25}).
Such cat-like oscillator states can be subsequently purified (separated from the qubits) and their parameters (mutual proportions and dynamical relations of individual wavepackets) can be engineered with respect to potential applications.
These curious objects and  the way of their preparation represent an instructive and amusing demonstration of complex behavior generated in one of the simplest quantum systems.

We use a generalized Rabi model, which describes the spin of length ${j=N/2}$ with ${N=1,2,3,\dots}$ coupled to a~quantum oscillator \cite{Rabi36,Dick54,Jayn63} (see also \cite{Stra21,Stra24}).
The spin\,---\,its ${(2j\!+\!1)}$-dimensional Hilbert space\,---\,can be realized via an ensemble of $N$ qubits restricted to the subspace of fully exchange-symmetric states.
This system can be experimentally implemented with the aid of trapped ions, macroscopic mechanical oscillators or superconducting circuits, see, e.g., Refs.\,\cite{Diaz19,Stef19,Camp20,Stas20,Hast21b,Chen21,Eick22,Siva22,Wang23,Pan23}. 
Denoting by $\hat{b}^{\dag}$ and $\hat{b}$ the creation and annihilation operators of the bosonic oscillator quanta and by ${\hat{J}_{\pm}=\hat{J}_{x}\pm i\hat{J}_{y}}$, ${\hat{J}_{0}=\hat{J}_{z}}$ the angular-momentum operators of the spin (for qubits ${\hat{\boldsymbol{J}}=\frac{1}{2}\sum_{i=1}^N\hat{\boldsymbol{\sigma}}_i}$, where $\hat{\boldsymbol{\sigma}}_i$ is the vector of Pauli matrices acting in the $i$th-qubit space), the Hamiltonian reads as
\begin{equation}
\begin{array}{rl}
\displaystyle{\frac{\hat{H}(\lambda)}{\omega}}=\hat{b}^{\dag}\hat{b}+R(\hat{J}_0\!+\!j)+\lambda\sqrt{\frac{R}{8j}} 
\bigl[ & \!\!\!  (1\!+\!\delta)(\hat{b}^{\dag}\hat{J}_{-}\!+\!\hat{b}\hat{J}_{+})
\\ 
+ & \!\!\!  (1\!-\!\delta)(\hat{b}^{\dag}\hat{J}_{+}\!+\!\hat{b}\hat{J}_{-})\bigr],
\end{array}
\label{Ham}
\end{equation}
where $\omega$ and $R\omega$, respectively, are energies of the elementary oscillator and spin excitations, and $\lambda$ is a dimensionless spin-oscillator coupling strength.
The character of the spin-oscillator interaction is varied by parameter $\delta$ between the Janes-Cummings (${\delta=+1}$), Dicke (${\delta=0}$) and anti-Janes-Cummings (${\delta=-1}$) regimes \cite{Dick54,Jayn63}.
Note that the dependence of $\hat{H}$ on $\delta$ and $R$ is implicit as these parameters stay fixed in the protocol described below.
In the following, we set ${\hbar=1}$ and express the energy $E$ and time $t$ in units ${\epsilon=2jR\omega}$ and ${\tau=1/\epsilon}$, respectively.

\begin{figure*}[!htbp]
    \centering
    \includegraphics[width=\linewidth]{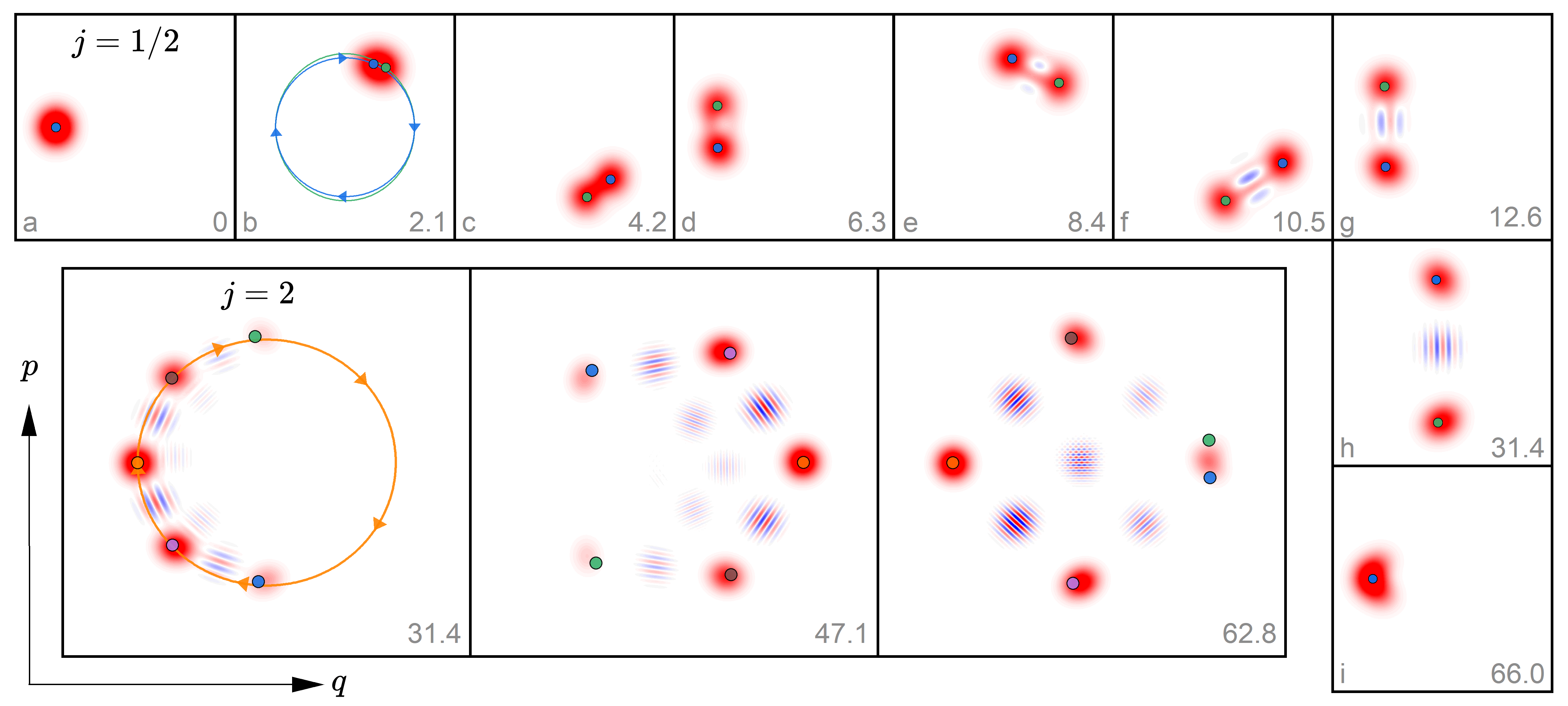}
    \caption{
The evolving oscillator Wigner functions $W_{\rm osc}(q,p,t)$ (with red/blue indicating positive/negative values) for systems with spins ${j=\frac{1}{2}}$ and $2$ after the quantum quench from ${\lambda_{\rm in}=1.5}$ to ${\lambda_{\rm fi}=-0.283}$, with ${\delta=0.5}$ and ${R=20}$.
The evolution starts at ${t=0}$ from a single wavepacket in the leftmost (${q<0}$) position. 
Times $t/\tau$ of all snapshots are indicated within the panels (for ${j=\frac{1}{2}}$ they are also marked in Fig.\,\ref{Puri}).
Color bullets and the oval curves indicate classical trajectories $(q(t),p(t))$.
The displayed phase-space domain in all cases is $q,p\in[-1.5,+1.5]$.
    }
    \label{Wig}
\end{figure*}

We assume that the ratio~$R$ between energies of the spin and oscillator elementary excitations satisfies the condition ${R\gg 1}$, which leads to semiclassical behavior of the oscillator as the average numbers of phonons in most of the system states are very large. 
This effective size parameter \cite{Cejn21} also enters the interaction term to ensure its proportional scaling with respect to the free Hamiltonian \cite{Stra21}.
In the limit ${R\to\infty}$, the system develops critical points at ${\lambda=\pm 1}$, where the ground state transforms nonanalytically between the normal (${|\lambda|<1}$) and superradiant (${|\lambda|>1}$) phases, which  is connected with spontaneous breaking of the conserved parity $\hat{\Pi}=(-1)^{\hat{b}^{\dag}\hat{b}+\hat{J}_0+j}$.
This ground-state quantum phase transition leads to the occurrence of nearly degenerate parity doublets in the low-energy spectrum for ${|\lambda|>1}$ and is accompanied by a pattern of excited-state singularities \cite{Cejn21,Stra21}.

Our method of creating multicomponent Schr{\"o}dinger-cat states relies on the quantum quench protocol, in which the spin-oscillator system is prepared nearly in the ground state ${\ket{\Psi_{\rm in}}\approx\ket{\Psi_0(\lambda_{\rm in})}}$ of Hamiltonian ${\hat{H}_{\rm in}\equiv\hat{H}(\lambda_{\rm in})}$ (more precisely, $\ket{\Psi_{\rm in}}$ coincides with a particular superposition of the lowest energy parity doublet in the superradiant phase) and then the coupling strength is suddenly switched to another value $\lambda_{\rm fi}$, yielding Hamiltonian ${\hat{H}_{\rm fi}\equiv\hat{H}(\lambda_{\rm fi})}$.
What happens is illustrated in Fig.\,\ref{Wig}.
It shows two examples of the Wigner function $W_{\rm osc}(q,p,t)$ \cite{Hill84,Polk10} associated with the density operator ${\hat{\rho}_{\rm osc}(t)={\rm tr}_{{\rm sp}}\ket{\Psi(t)}\bra{\Psi(t)}}$  of the oscillator subsystem evolving after the quench with given parameters.
Here ${\ket{\Psi(t)}=e^{-i\hat{H}_{\rm fi}t}\ket{\Psi_{\rm in}}}$ and ${\rm tr}_{{\rm sp}}$ is the partial trace over the spin Hilbert space.
The figure shows only a few snapshots of the after-quench dynamics. 
A much clearer picture can be obtained from the attached video files \cite{supp}.
We also note that the textual Supplemental Material \cite{supp} presents results for a wider range of model parameters, while the values used here serve just as examples.  

In general, for the system with a given $j$ we observe ${n=2j\!+\!1=N\!+\!1}$ wavepackets moving in the coordinate ($q$) $\times$ momentum ($p$) phase space of the oscillator.
The trajectory curves of individual wavepackets are very close, though not identical, but the speeds differ substantially.
From time to time various pairs of wavepackets cross each other within short intervals of overlap.
We stress that individual wavepackets represent nearly classical states of the oscillator, but their observed combination is a highly nonclassical and coherent quantum state.
The semiclassicality is indicated by the agreement of wavepacket trajectories with the classical predictions (which, as noted above, follows from the large value of~$R$), while the quantum coherence is manifested by the ripples of positive and negative values of $W_{\rm osc}(q,p,t)$ between each pair of wavepackets.

\begin{figure*}[!htbp]
    \centering
    \includegraphics[width=\linewidth]{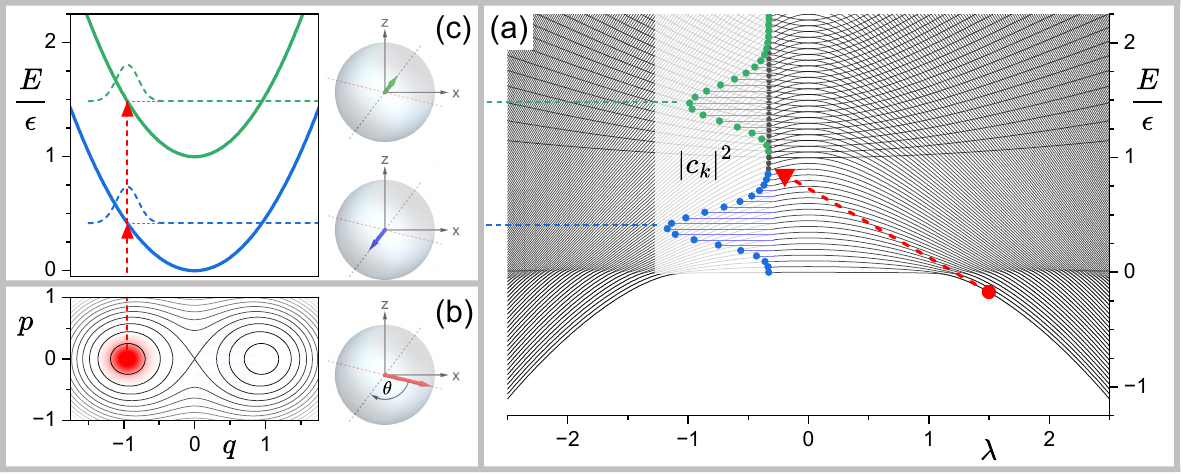}
\caption{(a) Spectrum (black lines) of Hamiltonian \eqref{Ham} and representation of the quench from Fig.\,\ref{Wig} for ${j=\frac{1}{2}}$. 
The final energy distribution $|c_k|^2$ is split to blue and green parts corresponding to ${m=-\frac{1}{2}}$ and ${+\frac{1}{2}}$, respectively.
(b)~Oscillator Wigner function (red blob) on the energy contours of $\hat{H}_{\rm osc}^{(-1/2)}$  (gray lines) and the spin orientation on the Bloch sphere at $\lambda_{\rm in}$.
(c)~Cuts of the oscillator effective Hamiltonians $\hat{H}_{\rm osc}^{(\pm 1/2)}$ (blue and green curves) and the spin orientation at $\lambda_{\rm fi}$.
    }
    \label{Schema}
\end{figure*}
 
The explanation of this behavior follows from the conversion of Hamiltonian \eqref{Ham} into the form 
\begin{equation}
\frac{\hat{H}(\lambda)}{\omega}=Rj+\hat{b}^{\dag}\hat{b}+\hat{\boldsymbol{B}}(\lambda)\cdot\hat{\boldsymbol{J}}=\hat{b}^{\dag}\hat{b}+|\hat{\boldsymbol{B}}(\lambda)|\ \hat{\boldsymbol{n}}(\lambda)\cdot\hat{\boldsymbol{J}},
\label{HamBa}
\end{equation}
where the vector operator ${\hat{\boldsymbol{J}}\equiv(\hat{J}_x,\hat{J}_y,\hat{J}_z)}$ acts in the spin subspace of the full Hilbert space, while vector operators $\hat{\boldsymbol{B}}$ and ${\hat{\boldsymbol{n}}=\hat{\boldsymbol{B}}/|\hat{\boldsymbol{B}}|}$ are expressed through $\hat{b}^{\dag}$ and~$\hat{b}$ defined in the oscillator subspace \cite{Bast17}.
Introducing coordinate and momentum operators ${\hat{q}=(\hat{b}^{\dag}\!+\!\hat{b})/\sqrt{4jR}}$ and ${\hat{p}={\rm i}(\hat{b}^{\dag}\!-\!\hat{b})/\sqrt{4jR}}$, we get ${\hat{\boldsymbol{B}}=R(\sqrt{2}\lambda\hat{q},-\sqrt{2}\lambda\delta\hat{p},1)}$ and ${|\hat{\boldsymbol{B}}|=R\sqrt{2\lambda^2(\hat{q}^2\!+\!\delta^2\hat{p}^2)\!+\!1}}$.
When approaching the limit ${R\to\infty}$, the commutator ${[\hat{q},\hat{p}]={\rm i}/2jR}$ vanishes and the observable $\hat{\boldsymbol{n}}\cdot\hat{\boldsymbol{J}}$ becomes an approximate integral of motions.
For semiclassical (narrow) wavepackets both $\hat{q}$ and~$\hat{p}$ can be replaced by their averages $\bar{q}$ and $\bar{p}$, so the operators $\hat{\boldsymbol{B}}(\lambda)$ and $\hat{\boldsymbol{n}}(\lambda)$ become ordinary functions $\boldsymbol{B}(\bar{q},\bar{p},\lambda)$ and $\boldsymbol{n}(\bar{q},\bar{p},\lambda)$, and $\hat{\boldsymbol{n}}\cdot\hat{\boldsymbol{J}}$ turns into $\hat{J}_{\boldsymbol{n}}$, which is the spin projection in the direction of the unit vector~$\boldsymbol{n}$.
This leads to full correlation of the spin and oscillator dynamics, when the spin state evolves in a way determined by the motion of the oscillator wavepacket in $q\times p$.
In particular, for a~given value ${m=-j,\dots,+j}$ of~$\hat{J}_{\boldsymbol{n}}$ the oscillator evolves by an effective Hamiltonian 
\begin{equation}
{\frac{\hat{H}_{\rm osc}^{(m)}(\lambda)}{\omega}=\hat{b}^{\dag}\hat{b}+|\hat{\boldsymbol{B}}(\lambda)|\, m},
\label{Heff}
\end{equation}
while the spin retains projection $m$ to the moving direction ${\boldsymbol{n}(t)=\boldsymbol{n}(\bar{q}(t),\bar{p}(t),\lambda)}$, where $\bar{q}(t)$ and $\bar{p}(t)$ stand for the evolving oscillator coordinate-momentum averages. 

The initial state ${\ket{\Psi_{\rm in}}\equiv\ket{\Psi(0)}}$ has an approximately factorized form ${\ket{\Psi_{\rm in}}\approx\ket{\psi_{\rm in}}\otimes\ket{{-j};{\boldsymbol{n}_{\rm in}}}}$, where $\ket{\psi_{\rm in}}$ is a~displaced quasi-Gaussian state of the oscillator and $\ket{{-j};{\boldsymbol{n}_{\rm in}}}$ denotes the spin state with the minimal projection in the direction ${\boldsymbol{n}_{\rm in}\equiv\boldsymbol{n}(\bar{q}_{\rm in},\bar{p}_{\rm in},\lambda_{\rm in})}$ corresponding to the initial ground-state averages ${\bar{q}_{\rm in}=\matr{\psi_{\rm in}}{\hat{q}}{\psi_{\rm in}}}$ and ${\bar{p}_{\rm in}=\matr{\psi_{\rm in}}{\hat{p}}{\psi_{\rm in}}}$.
This state cast in terms of the new spin direction ${\boldsymbol{n}'_{\rm in}=\boldsymbol{n}(\bar{q}_{\rm in},\bar{p}_{\rm in},\lambda_{\rm fi})}$, corresponding to  the final strength $\lambda_{\rm fi}$, reads as
\begin{equation}
\ket{\Psi(0)}\approx\ket{\psi_{\rm in}}\otimes\biggl[\,\sum_{m=-j}^{+j}\!\underbrace{\scal{m;{\boldsymbol{n}'_{\rm in}}}{-\!j;{\boldsymbol{n}_{\rm in}}}}_{\alpha^{(m)}}\ket{m;{\boldsymbol{n}'_{\rm in}}}\biggr].
\label{psina}
\end{equation}
The evolution of this state is then given by
\begin{equation}
\ket{\Psi(t)}\approx\!\!\sum_{m=-j}^{+j}\!\!\alpha^{(m)}\underbrace{\bigl[e^{-i\hat{H}^{(m)}_{\rm osc}(\lambda_{\rm fi})t}\ket{\psi_{\rm in}}\bigr]}_{\ket{\psi^{(m)}(t)}}\!\otimes\ket{m;\boldsymbol{n}^{(m)}(t)},
\label{kockovina}
\end{equation}
where the direction ${\boldsymbol{n}^{(m)}(t)=\boldsymbol{n}(\bar{q}^{(m)}(t),\bar{p}^{(m)}(t),\lambda_{\rm fi})}$ is determined from moving coordinate and momentum ($\hat{x}\equiv\hat{q},\hat{p}$) averages ${\bar{x}^{(m)}(t)=\matr{\psi^{(m)}(t)}{\hat{x}}{\psi^{(m)}(t)}}$.
Since the effective oscillator Hamiltonians \eqref{Heff} for various projections $m$ differ from each other, the initial wavepacket $\ket{\psi_{\rm in}}$ splits into ${n=2j\!+\!1}$ components $\ket{\psi^{(m)}(t)}$ moving in a different way, which can be seen in the phase-space representation $W_{\rm osc}(q,p,t)$ of the oscillator state.
The weights of these components are given by $|\alpha^{(m)}|^2$ and their partial coherence follows from the fact that the accompanying spin states $\ket{m;\boldsymbol{n}^{(m)}(t)}$ are not mutually orthogonal.

The above analysis is illustrated by an example in Fig.\,\ref{Schema}.
Here, the system with ${j=\frac{1}{2}}$  is prepared  in the initial state of the quench from Fig.\,\ref{Wig}.
This state (a particular superposition of the quasidegenerate lowest energy parity doublet in the superradiant phase) is represented by the bullet in the energy spectrum $\{E_k(\lambda)\}$ in panel~(a) and by the factorized oscillator-spin state in panel~(b), where the oscillator Wigner function is located in one of two nearly degenerate minima of the classical $q\times p$ representation of $\hat{H}_{\rm osc}^{(-1/2)}(\lambda_{\rm in})$ while the spin points along the direction $\boldsymbol{n}_{\rm in}$ on the Bloch sphere.
Parameter $\lambda$ is then switched to a negative $\lambda_{\rm fi}$ in the normal phase.
The arrow in panel~(a) ends at the final energy average $\matr{\Psi_{\rm in}}{\hat{H}_{\rm fi}}{\Psi_{\rm in}}$ and the histogram depicts the final energy distribution ${|c_k|^2=|\scal{E_k(\lambda_{\rm fi})}{\Psi_{\rm in}}|^2}$. 
This distribution has two peaks centered at the mean energies of the initial state with respect to the new effective Hamiltonians $\hat{H}_{\rm osc}^{(-1/2)}(\lambda_{\rm fi})$ and $\hat{H}_{\rm osc}^{(+1/2)}(\lambda_{\rm fi})$, corresponnding to the decomposition of the initial spin state to both projections along the new direction $\boldsymbol{n}'_{\rm in}$, that forms angle $\theta$ with the original direction $\boldsymbol{n}_{\rm in}$.
The integral probability in both peaks is equal to the above-defined values $|\alpha^{(\pm 1/2)}|^2$.
The ${p=0}$ cuts of both $\hat{H}_{\rm osc}^{(\pm 1/2)}(\lambda_{\rm fi})$ and the new spin direction are shown in panel~(c).
This indicates the starting point of the after-quench evolution, when the initial state splits into the ${m=\pm\frac{1}{2}}$ parts, each of them composed of the oscillator wavepacket moving in the respective effective Hamiltonian $\hat{H}_{\rm osc}^{(\pm 1/2)}(\lambda_{\rm fi})$ and the spin keeping projection $m$  to the evolving direction $\boldsymbol{n}^{(\pm 1/2)}(t)$ determined by the respective oscillator state. 

\begin{figure}[!tbp]
    \centering
    \includegraphics[width=\linewidth]{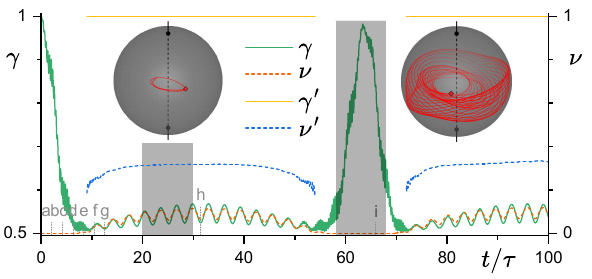}
        \caption{Purity $\gamma(t)$ and negativity $\nu(t)$ after the quench from Fig.\,\ref{Wig} for ${j=\frac{1}{2}}$. 
        Inserted Bloch-sphere diagrams depict motions of the average spin $\langle{\boldsymbol J}(t)\rangle$ in two time intervals indicated by the gray zones.
        For some time intervals we also show the purity ${\gamma'=1}$ and negativity $\nu'(t)$ resulting from the ${\boldsymbol{n}'\bot\,\boldsymbol{n}^{(\pm\frac{1}{2})}(t)}$  spin measurement at time $t$.
        }
    \label{Puri}
\end{figure}

In Fig.\,\ref{Puri} we show the purity and the negativity of the oscillator state in a system with ${j=\frac{1}{2}}$ after the quench from Fig.\,\ref{Wig}.
The purity ${\gamma(t)={\rm tr}\,\hat{\rho}_{\rm osc}^2(t)}$ expresses the spin-oscillator entanglement \cite{Niel00}, yielding values between ${\gamma_{\rm min}=\frac{1}{2j+1}}$ (for the maximally mixed oscillator state, i.e., maximally entangled total state $\ket{\Psi(t)}$) and ${\gamma_{\rm max}=1}$ (for a~pure oscillator state, i.e., fully factorized total state). 
The negativity ${\nu(t)=\frac{1}{2}\int [|W_{\rm osc}(q,p,t)|\!-\!W_{\rm osc}(q,p,t)]dq dp}$ measures the negative volume of the oscillator Wigner function \cite{Kenf04}, quantifying the quantum coherence indicated by the ripples between individual wavepackets. 
We observe that the purity for ${j=\frac{1}{2}}$  oscillates at the level ${\gamma\!-\!\gamma_{\rm min}\approx 0.1(\gamma_{\rm max}\!-\!\gamma_{\rm min})}$ for most of time.
For higher spins, the oscillations occur around higher values, e.g., ${\gamma\!-\!\gamma_{\rm min}\approx 0.25(\gamma_{\rm max}\!-\!\gamma_{\rm min})}$ for ${j=2}$ \cite{supp}.
On the other hand, the purity reaches the maximal value $\gamma_{\rm max}$ within the time intervals when the negativity $\nu$ gets close to zero.
Indeed, in these intervals both ${m=\pm\frac{1}{2}}$ wavepackets meet and overlap, so the spin-oscillator state becomes approximately factorized.
As a consequence of the constructive addition of individual spin components in the overlap intervals, the average spin ${\langle{\boldsymbol J}(t)\rangle\equiv\matr{\Psi(t)}{\hat{{\boldsymbol J}}}{\Psi(t)}}$ grows from the interior of the Bloch sphere with radius~$j$ to its surface; see the second inset in Fig.\,\ref{Puri}.
Simultaneously, the average spin exhibits very rapid precessional motions along the uniquely defined direction of $\boldsymbol{B}(\bar{q},\bar{p},\lambda)$.

\begin{figure}[!htbp]
    \centering
    \includegraphics[width=\linewidth]{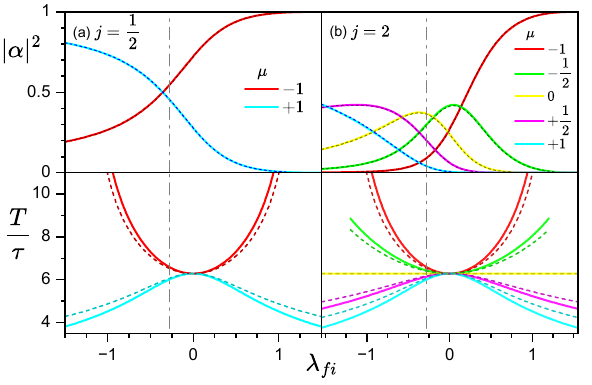}
        \caption{Cat-state parameters for ${\lambda_{\rm in}=1.5}$ and variable~$\lambda_{\rm fi}$ with (a) ${j=\frac{1}{2}}$ and (b) ${j=2}$. 
        Upper panels: Weights $|\alpha^{(m)}|^2$ of individual wavepackets from Eq.\,\eqref{psina}. 
        Solid and dashed curves (practically coinciding) are calculated, respectively, from $d$-functions \cite{supp} and by summing $|c_k|^2$ in the corresponding peaks, cf. Fig.\,\ref{Schema}(a).
        Lower panels: Periods of classical orbits for ${\mu=\frac{m}{j}=-1,\cdots,+1}$ and two values $\delta=0.5$ (solid lines) and $\delta=0$ (dashed lines).
        The vertical gray line indicates the quench used in the previous figures.
        We set $R=50$.
        }
    \label{Tuni}
\end{figure}

Figure~\ref{Tuni} shows the weights of the ${2j+1}$ wavepackets (for ${j=\frac{1}{2}}$ and ${j=2}$) and the periods of their motions as functions of the quench parameter $\lambda_{\rm fi}$ for two values of the Hamiltonian parameter $\delta$.
We find that adjustments of the quench length and $\delta$ enable us to create the multicomponent cat-like states in a large variety of proportions and dynamical relations. 
This is also demonstrated in the Supplemental Material and appended animations \cite{supp}.
However, in spite of these positive characteristics, our method suffers from an apparent drawback\,---\,the limited purity of the cat-like states produced (Fig.\,\ref{Puri}).
This results from the high degree of entanglement between the spin and oscillator subsystems, which is an inherent ingredient of the procedure. 

\begin{figure}[!tbp]
    \centering
    \includegraphics[width=\linewidth]{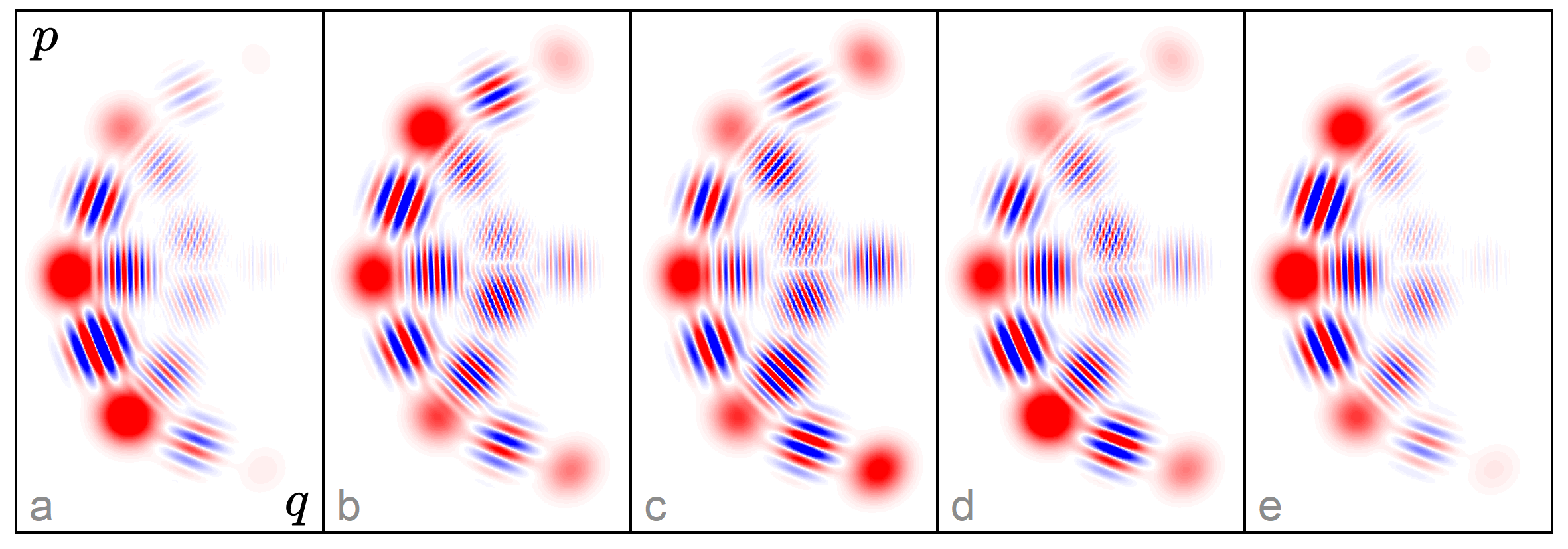}
\caption{Wigner function $W_{\rm osc}(q,p,t)$ corresponding to the ${j=2}$ case at ${t=31.4}$ from Fig.\,\ref{Wig} after the purification measurement on the spin subsystem.
Panels from a to e correspond to measured spin projections $m'$ from $-2$ to $+2$ in the direction ${(\vartheta,\varphi)=(1.54,3.08)}$ on the Bloch sphere.
    }
    \label{Purif}
\end{figure}

To eliminate this deficiency, we propose a simple purification technique that eventually breaks the spin-oscillator entanglement without destroying the cat state character.
It consists of performing a suitable measurement on the spin subsystem.
Indeed, the spin measurement on the entangled state \eqref{kockovina} at time~$t$, which results in finding the spin projection~$m'$  in the direction~$\boldsymbol{n}'$, creates a factorized state
\vspace{-3mm}
\begin{eqnarray}
\ket{\Psi'(t)}\approx\!\overbrace{\biggl[\mathcal{N}\!\!\sum_{m=-j}^{+j}\!\!\alpha^{(m)}\scal{m';\boldsymbol{n}'}{m;\boldsymbol{n}^{(m)}(t)}\ket{\psi^{(m)}(t)}\biggr]}^{\ket{\psi'(t)}}
\qquad\nonumber\\
\otimes\ \ket{m';\boldsymbol{n}'}.\qquad
\label{purecat}
\end{eqnarray} 
Here ${\ket{\psi'(t)}=\sum_m \alpha^{(m)}\,\!'\ket{\psi^{(m)}(t)}}$ is a pure state of the oscillator (purity ${\gamma'=1}$) with new coefficients defined in Eq.\,\eqref{purecat}, where~$\mathcal{N}$ is a normalization coefficient.
For a~suitable choice of $\boldsymbol{n}'$ and $m'$, this state retains the original character of the multicomponent cat state.
The optimal solution can be found for ${j=\frac{1}{2}}$,  when the measurement direction $\boldsymbol{n}'$ perpendicular to both directions $\boldsymbol{n}^{(\pm\frac{1}{2})}(t)$ conserves the proportions of the two cat-state components for both ${m'=\pm\frac{1}{2}}$ measurement outcomes (i.e., it yields a pure state with ${|\alpha^{(m)}\,\!'|^2=|{\alpha^{(m)}}|^2}$ for ${m=\pm\frac{1}{2}}$).
The negativity $\nu'(t)$ of such purified states was also drawn in Fig.\,\ref{Puri}.
For ${j>\frac{1}{2}}$, the direction $\boldsymbol{n}'$  perpendicular to all $\boldsymbol{n}^{(m)}(t)$ cannot, in general, be found, so the measurement is likely to change proportions of the multiple cat components.
Nevertheless, spin measurements in a large variety of directions, with various outcomes $m'$, lead to rather satisfactory results, as exemplified in Fig.\,\ref{Purif} for ${j=2}$.  
The Wigner functions of the purified 5-component cat state in this figure should be compared with the corresponding mixed state from Fig.\,\ref{Wig}.
A~more detailed discussion of the choice of $\boldsymbol{n}'$ and $m'$ can be found in the Supplemental Material \cite{supp}.

The above-presented method may be used to create specific cat-like states tailored to some particular applications.
The state preparation protocol consists of the following steps:
(i) formation of the initial state $\ket{\Psi_{\rm in}}$, a superposition of the superradiant ground-state parity doublet at ${\lambda=\lambda_{\rm in}}$,
(ii) realization of the prescribed quench $\lambda_{\rm in}\to\lambda_{\rm fi}$,
(iii) spontaneous evolution of the coupled system with ${\lambda=\lambda_{\rm fi}}$ for time $t$,
(iv) measurement of the spin subsystem in a suitable direction $\boldsymbol{n}'$ and filtering out a~suitable outcome $m'$,
(v) turning off the spin-oscillator interaction: ${\lambda\to 0}$.
Since then all components of the purified state $\ket{\psi'(t)}$ just periodically circulate in the intact harmonic oscillator potential and are stored for further application.

In conclusion, we have described a method of creating ${(2j\!+\!1)}$-component cat-like states in a semiclassical oscillator coupled to spin of size $j$ (to a~set of ${N=2j}$ qubits in an exchange-symmetric state).
The protocol, which employs quantum quench dynamics of the spin-oscillator system in the coupling strength $\lambda$ and a subsequent measurement of the spin subsystem, yields the cat state with unit purity and allows for tuning its parameters. 
In particular, it is capable of creating various motion configurations of a variable number of wave packets in different proportions.
The proposed method can be potentially used in quantum information and sensing procedures involving qudits \cite{Wang20,Len22}.
In a wider perspective, the above-outlined technique for controlling a continuous quantum system via its coupling to a discrete system represents an important general scheme that offers a vast variety of potential applications.

{\em Acknowledgments---\/}Quantum dynamics was computed with the aid of the QuantumOptics.jl package~\cite{Julia}.
We acknowledge financial support from the Czech Science Foundation under Project No.\,25-16056S.

{\em Data availability---\/}The data that support the above findings can be obtained by openly available codes~\cite{data}.

\end{document}


\title{Supplemental material: Creating multicomponent Schrödinger cat states in a coupled qubit-oscillator system}
\date{\today}
\author{Pavel Str\'ansk\'y}
\email{pavel.stransky@matfyz.cuni.cz}
\affiliation{Institute of Particle and Nuclear Physics, Faculty of Mathematics and Physics, Charles University, V Hole\v{s}ovi\v{c}k\'ach 2, 18000 Prague, Czechia}
\author{Pavel Cejnar}
\email{pavel.cejnar@matfyz.cuni.cz}
\affiliation{Institute of Particle and Nuclear Physics, Faculty of Mathematics and Physics, Charles University, V Hole\v{s}ovi\v{c}k\'ach 2, 18000 Prague, Czechia}

\begin{abstract}
In this supplemental text, we provide additional details on the extended Rabi model, derivations of key formulas, and supplementary figures that support the findings presented in the main text of the paper in order to demonstrate the range of possible after-quench outcomes.
\end{abstract}

\maketitle

\section{Model}
	We consider an extended version of the Rabi model (also called the generalized~\cite{Hioe1973,Tomka2014}, asymmetric~\cite{Shen2014} or anisotropic~\cite{Xie2014,Zhang2017,Liu2017,Shen2017,Buijsman2017,Das2023,Zhu2024} Rabi or Dicke model) with the Hamiltonian parametrized as
	\begin{align}
		\label{eq:H}
		\operator{H}&=
			\omega\operatorconjugate{b}\operator{b}+\omega R\operator{J}_{z}
			+2\lambda\omega\sqrt{\frac{R}{8j}}\lambda\left[\left(\operatorconjugate{b}+\operator{b}\right)\operator{J}_{x}-\im\delta\left(\operatorconjugate{b}-\operator{b}\right)\operator{J}_{y}\right]\nonumber\\
			&=\omega\operatorconjugate{b}\operator{b}+\omega R\operator{J}_{z}
			+\lambda\omega\sqrt{\frac{R}{8j}}\left[\left(1+\delta\right)\left(\operatorconjugate{b}\operator{J}_{-}+\operator{b}\operator{J}_{+}\right)+\left(1-\delta\right)\left(\operatorconjugate{b}\operator{J}_{+}+\operator{b}\operator{J}_{-}\right)\right],
	\end{align}
	where $\vectoroperator{J}=\left(\operator{J}_{x},\operator{J}_{y},\operator{J}_{z}\right)$ is the quasispin operator of size $j$, $\operator{b},\operatorconjugate{b}$ are the field (oscillator) annihilation and creation operators,
	$\omega$ is the field frequency,
	$R\gg1$ is the detuning between the quasispin and the oscillator, which also plays the role of the size parameter of the system~\cite{Stransky2021a},
	$\lambda$ is the quasispin-field interaction strength,
	and $\delta$ interpolates between the ($\delta=1$) Jaynes (Tavis)-Cummings regime, ($\delta=0$) Rabi (Dicke) regime and ($\delta=-1$) anti-Jaynes (anti-Tavis)-Cummings regime.

	The critical coupling---Quantum Phase Transition (QPT)---between the normal ($\abs{\lambda}<1$) and deformed ($\abs{\lambda}>1$) phases occurs at $\lambda_{c}=\pm1$, independently of $\delta$ and $j$.

	The whole Hamiltonian can be rescaled and shifted so that all its terms are dimensionless and the ground-state energy of the noninteracting configuration $E_{0}(\lambda=0)$ vanishes:
	\begin{align}
		\label{eq:hoperator}
		\operator{h}
			&\equiv\frac{1}{2j\omega R}\left(\operator{H}+2j\omega R\right)\nonumber\\
			&=\frac{1}{2jR}\operatorconjugate{b}\operator{b}+\frac{1}{2}\left(\operator{j}_{z}+1\right)
				+\lambda\frac{1}{4j}\sqrt{\frac{1}{2jR}}\left[\left(1+\delta\right)\left(\operatorconjugate{b}\operator{j}_{-}+\operator{b}\operator{j}_{+}\right)+\left(1-\delta\right)\left(\operatorconjugate{b}\operator{j}_{+}+\operator{b}\operator{j}_{-}\right)\right],
	\end{align}
	where $\operator{j}_{\bullet}=\operator{J}_{\bullet}/j$.
	After introducing the coordinate and momentum by substituting $\operator{b}=\sqrt{jR}\left(\operator{q}+\im\operator{p}\right)$, the Hamiltonian becomes
	\begin{equation}
		\label{eq:h}		
			\operator{h}
				=\frac{1}{2}\left(\operator{p}^{2}+\operator{q}^{2}\right)-\frac{1}{jR}+\frac{1}{2}+\vectoroperator{b}\cdot\vectoroperator{j},
	\end{equation}
	where the rescaled effective magnetic field, independent of the size parameter $R$, is
	\begin{equation}
		\label{eq:B}
		\vectoroperator{b}\equiv\frac{\vectoroperator{B}}{2R}=\left(\frac{\lambda}{\sqrt{2}}\operator{q},-\frac{\lambda}{\sqrt{2}}\delta\operator{p},\frac{1}{2}\right)=\abs{\vectoroperator{b}}\vectoroperator{n}.
 	\end{equation}
	$\vectoroperator{B}$ is the effective magnetic field defined in the main text and $\vectoroperator{n}$ is the corresponding unit vector.

	In the classical limit $R\rightarrow\infty$, the classical Hamiltonian reads as
	\begin{equation}
		\label{eq:hcl}			
		\h{\mu}(p,q,\lambda)=\frac{1}{2}\left(p^{2}+q^{2}\right)+\frac{1}{2}+\mu\sqrt{\frac{{\lambda}^2}{2}\left(q^{2}+\delta^{2}p^{2}\right)+\frac{1}{4}},
	\end{equation}
	where $\mu=m/j\in[-1,1]$ and $m=-j,\dotsc,j$ is the quasispin projection in the inner frame whose orientation is given by vector $\vector{B}$ (for more discussion see Ref.~\cite{Stransky2021a} and the main text).
	
\section{Quantum quench}
	The main conclusions of this work are based on the analysis of the dynamics after performing a quantum quench $\li\rightarrow\lf$ from a parity violating ground state of the deformed phase, $\li>1$, into $\lf<\li$.
	The initial energy calculated semiclassically from the global minimum of the Hamiltonian~\eqref{eq:hcl} placed at
	\begin{equation}
		q_{\mathrm{min}}(\lambda)=\pm\sqrt{\frac{1}{2}\left(\lambda^{2}-\frac{1}{\lambda^{2}}\right)},\qquad p_{\mathrm{min}}(\lambda)=0
	\end{equation}
	is
	\begin{equation}
		e_{\mathrm{min}}(\lambda)\equiv\h{-1}\left(p_{\mathrm{min}},q_{\mathrm{min}},\lambda\right)=-\frac{1}{4}\left(\lambda^{2}+\frac{1}{\lambda^{2}}\right)+\frac{1}{2}.
	\end{equation}
	Note that the global minimum is always located in $\mu=-1$ subspace~\cite{BastarracheaMagnani2017} and both its energy and position are independent of $\delta$ and $j$.

	The after-quench energy expectation value can be determined by the semiclassical Hellmann-Feynman formula
	\begin{align}
		\ef(\lf,\li)
			&=\ei+\left(\lf-\li\right)\partialderivative{\h{-1}}{\lambda}\left(p_{\mathrm{min}},q_{\mathrm{min}}, \li\right)\nonumber\\
			&=\frac{1}{4}\left(\li^{2}-\frac{3}{\li^{2}}\right)-\frac{1}{2}\lf\left(\li-\frac{1}{\li^{3}}\right)+\frac{1}{2},
	\end{align}
	where $\ei\equiv e_{\mathrm{\min}}(\li)$.

	First, let us focus on the single-spin case $j=\frac{1}{2}$.
	The initial state is approximately the product state~\cite{Puebla2020}
	\begin{equation}
		\label{eq:InitialState}
		\ket{\Psi(0)}=\ket{\psi_{\mathrm{in}}}\ket{\downarrow_{\Bi}},
	\end{equation}
	where $\ket{\psi_{\mathrm{in}}}\equiv\ket{q_{\mathrm{min}}}$ is the coherent state of the oscillator at $q=q_{\mathrm{min}}, p=p_{\mathrm{min}}=0$,
	\begin{equation}
		\Bi=\vector{b}(p_{\mathrm{min}},q_{\mathrm{min}},\li)=\left(\frac{\li}{\sqrt{2}}q_{\mathrm{min}},0,\frac{1}{2}\right)
	\end{equation}
	is the effective magnetic field for state $\ket{\psi_{\mathrm{in}}}$ and the spin points against the direction of this field; note that in the notation of the main text, $\ket{\downarrow_{\Bi}}\equiv\ket{m=1/2;\vector{n}_\mathrm{in}=\Bi/\abs{\Bi}}$.

	Immediately after the quench, the state is the same. 
	Expanded in the basis of the Hamiltonian for $\lf$, it can be expressed as
	\begin{equation}
		\label{eq:Expansion1/2}
		\ket{\Psi(0)}=\a{-1}\ket{q_{\mathrm{min}}}\ket{\downarrow_{\Bf}}+\a{+1}\ket{q_{\mathrm{min}}}\ket{\uparrow_{\Bf}},
	\end{equation}
	with amplitudes
	\begin{subequations}
		\label{eq:alpha12}
		\begin{align}
			\a{-1}&=\cos{\frac{\theta}{2}}=\sqrt{\frac{1-\cos{\theta}}{2}},\\
			\a{+1}&=\e^{\im\phi}\sin{\frac{\theta}{2}}=\e^{\im\phi}\sqrt{\frac{1+\cos{\theta}}{2}},
		\end{align}				
	\end{subequations}
	where $\theta,\phi$ are the polar and azimuthal angles, respectively, of the effective magnetic field vector
	\begin{equation}
		\Bf=\left(\frac{\lf}{\sqrt{2}}q_{\mathrm{min}},0,\frac{1}{2}\right)		
	\end{equation}
	with respect to the coordinate frame in which the initial magnetic field $\Bi$ points along the $z$-axis. 
	The polar angle $\theta$ can be explicitly calculated as
	\begin{equation}
		\label{eq:costheta}
		\cos{\theta}=\frac{\Bi\cdot\Bf}{\abs{\Bi}\abs{\Bf}}=\frac{2\li\lf q_{\mathrm{min}}^{2}+1}{\sqrt{\left(2\li^{2}q_{\mathrm{min}}^{2}+1\right)\left(2\lf^{2}q_{\mathrm{min}}^{2}+1\right)}}=\frac{1+\lf\li\left(\li^{2}-\frac{1}{\li^{2}}\right)}{\li^{2}\sqrt{1+\lf^{2}\left(\li^{2}-\frac{1}{\li^{2}}\right)}},
	\end{equation}
	and the coordinate systems can be chosen so that the azimuthal angle is $\phi=0$.
	
	\begin{figure}[htbp!]
		\includegraphics[width=0.8\textwidth]{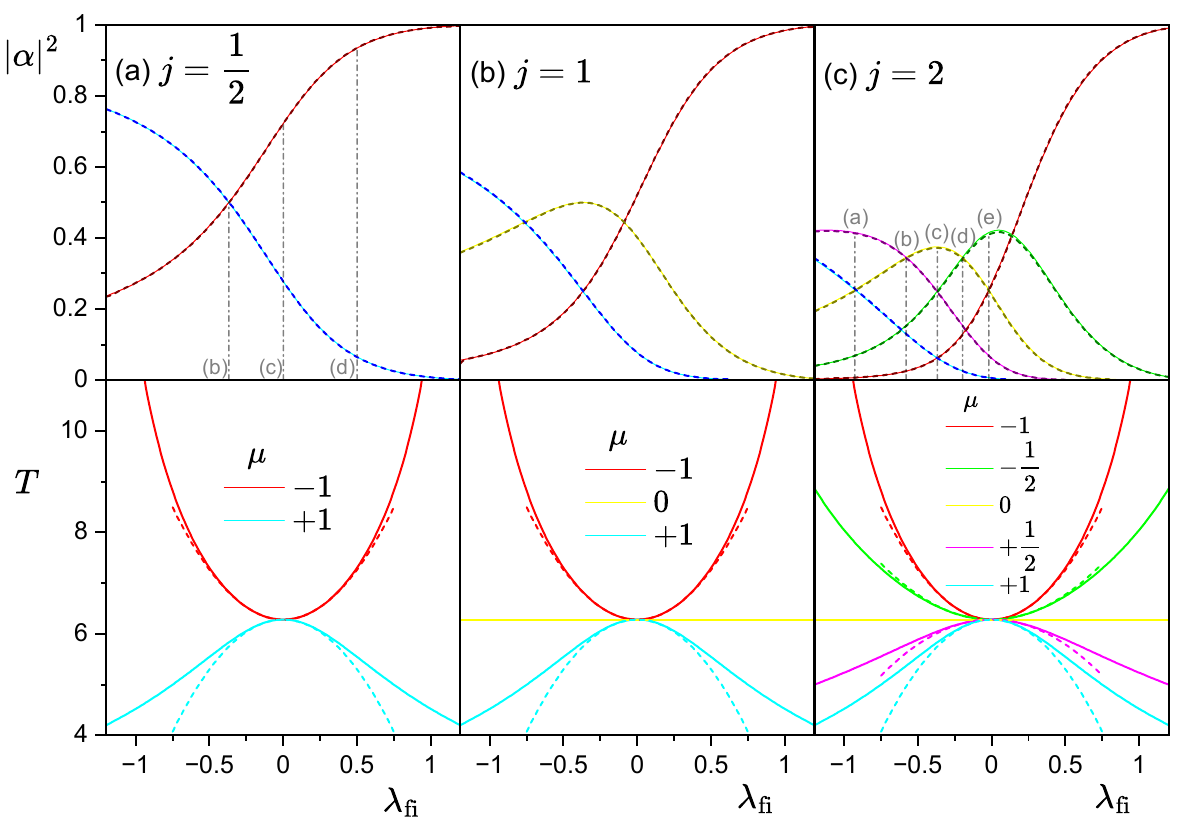}
		\caption{\emph{Top panels:} Squared amplitudes of the wavefunction contributions calculated from the Wigner $d$-functions (solid lines), see Eqs.~\eqref{eq:alpha12} and~\eqref{eq:alphagen}, and from the strength function~\ref{eq:SF} (dashed lines), for three values of $j$.
		Gray vertical dash-dot lines indicate the quenches shown in Figs.~\ref{fig:SF1} and~\ref{fig:SF4}.
		\emph{Bottom panels:} Frequencies of the classical trajectories in all available subspaces $\mu=m/j$, $m=-j,\dotsc,j$.
        Solid and dashed lines, respectively, show exact results and the weak-interaction approximation~\eqref{eq:Weak}.
		The initial coupling is $\li=1.5$ and the size parameter is $R=50$.
		}
		\label{fig:Amplitudes}
	\end{figure}

	The expansion coefficients $\a{-1},\a{+1}$ can also be determined entirely using semiclassical arguments.
	Final energy $\ef$ is composed of contributions from $\mu=\pm1$ subspaces with weights $\abs{\a{\pm1}}^{2}$.
	This leads to a set of equations
	\begin{align}
		\ef=&\abss{\a{-1}} \h{-1}(q_{\mathrm{min}},0,\lf)+\abss{\a{+1}} \h{+1}(q_{\mathrm{min}},0,\lf),\\
		&\abss{\a{-1}}+\abss{\a{+1}}=1,
	\end{align}
	whose solution gives exactly the same expressions as above, see Eq.~\eqref{eq:alpha12} and Fig.~\ref{fig:Amplitudes}(a).

	One can ask which quench gives balanced contributions to both subspaces $\mu=\pm1$ in the after-quench state, \ie which $\lf$ results in $\abs{\a{-1}}^{2}=\abs{\a{+1}}^{2}=1/2$.
	This happens when $\cos{\theta}=0$, corresponding to the after-quench coupling
	\begin{equation}
		\label{eq:SameMinMaxContribution}
		\lf^{(=)}\equiv\frac{\li}{1-\li^{4}}.
	\end{equation}
	The squared moduli of both $\a{-1}$ and $\a{+1}$ are shown by the solid lines in the first row of Fig.~\ref{fig:Amplitudes}(a) for $\li=1.5$.
	For this choice of the initial coupling, the moduli are equal when $\lf^{(=)}=-24/65\approx-0.369$.

	Another particular quench can be related to the presence of a stationary point in the classical Hamiltonian~\eqref{eq:hcl}.
	The Hamiltonian in subspace $\mu=-1$ and in the deformed quantum phase $\abs{\lambda_{c}}<\abs{\lambda}<\abs{\lambda_{c}/\delta}$ has a nondegenerate stationary point with index 1 (a saddle point).
	If a trajectory hits this point, \ie if $\h{-1}(0,q_{\mathrm{min}},\lf)=0$, its period diverges.
	Since
	\begin{equation}
		\h{-1}(0,q_{\mathrm{min}},\lf)=\frac{1}{4}\left(\li^{2}-\frac{1}{\li^{2}}\right)-\frac{1}{2}\sqrt{\lf^{2}\left(\li^{2}-\frac{1}{\li^{2}}\right)+1}+\frac{1}{2},
	\end{equation}
	this \emph{critical quench} is given by
	\begin{equation}
		\label{eq:CriticalQuench}
		\lf^{(c)}=\pm\sqrt{\frac{1}{4}\left(\li^{2}-\frac{1}{\li^{2}}\right)+1}.
	\end{equation}

	For $\li=3/2$, two possible critical quenches are $\lf^{(c)}=\pm\sqrt{209}/12\approx\pm1.205$.
	The evolution of the Wigner quasiprobability distribution after the quench with $\lf=-1.205$ is shown in the animation named \emph{Dos gotas de Rabi}~\cite{YouTubeBruder}.

	Last special quench discussed here is $\lf=-\li$, after which the part of the after-quench state in $\mu=-1$ subspace remains stationary in the initial minimum at $p=0,q=q_{\mathrm{min}}$, since this point is also a minimum of the after-quench Hamiltonian due to the $\h{\mu}(\lambda)=\h{\mu}(-\lambda)$ symmetry.
	The animation of this quench is shown as \emph{Rabiho rezonance dvou}~\cite{YouTubeBruder}.

	Let us now extend the discussion to an arbitrary spin $j$.
	The expansion~\eqref{eq:Expansion1/2} in the $\lf$ spin direction is given by the Wigner $d$-functions
	\begin{equation}
		\label{eq:alphagen}
		\ket{\Psi^{(j)}(0)}=\ket{q_{\mathrm{min}}}\otimes\left[\sum_{m=-j}^{j}d_{m,-j}^{(j)}(\theta)\ket{m}\right],
	\end{equation}
	where we use a shorthand notation $\ket{m}\equiv\ket{m;\vector{n}'_{\mathrm{in}}=\Bf/\abs{\Bf}}$.
	One can identify $\a{j,\mu}\equiv d^{(j)}_{\mu j,-j}(\theta)$, where angle $\theta$ is given by Eq.~\eqref{eq:costheta}.
	Examples for two values of $j>1/2$ are:
	\begin{itemize}
		\item $j=1$:
		\begin{equation}
			\ket{\Psi^{(1)}(0)}=\ket{q_{\mathrm{min}}}\left[\left(\cos{\frac{\theta}{2}}\right)^{2}\ket{-1}+\sqrt{2}\cos{\frac{\theta}{2}}\cos{\frac{\theta}{2}}\ket{0}+\left(\sin{\frac{\theta}{2}}\right)^{2}\ket{1}\right]
		\end{equation}

		\item $j=2$:
		\begin{align}
			\ket{\Psi^{(2)}(0)}
				&=\ket{q_{\mathrm{min}}}\Bigg[\left(\cos{\frac{\theta}{2}}\right)^{4}\ket{-2}+2\left(\cos{\frac{\theta}{2}}\right)^{3}\sin{\frac{\theta}{2}}\ket{-1}\nonumber\\
				&\quad+\sqrt{6}\left(\cos{\frac{\theta}{2}}\right)^{2}\left(\sin{\frac{\theta}{2}}\right)^{2}\ket{0}\\
				&\quad2\cos{\frac{\theta}{2}}\left(\sin{\frac{\theta}{2}}\right)^{3}\ket{+1}
				+\left(\sin{\frac{\theta}{2}}\right)^{4}\ket{+2}\Bigg]\nonumber
		\end{align}			
	\end{itemize}
	The corresponding squared moduli of the amplitudes $\abs{\a{j,\mu}}^{2}$ are shown in Fig.~\ref{fig:Amplitudes}(b,c).

	Note that 
	\begin{itemize}
		\item 
			Eq.~\eqref{eq:SameMinMaxContribution} for equal contributions from the lowest and highest subspaces $\mu=\pm1$ remains valid for any $j$; this equation even holds for any pair of subspaces with $+\mu$ and $-\mu$.
		\item
			The critical quench~\eqref{eq:CriticalQuench} also remains unchanged for general $j$.
			On top of that, for higher values of $j$, there exists more subspaces with unstable stationary points at $p=q=0$, see~\cite{BastarracheaMagnani2017,Stransky2021a}, so there will be more critical quenches.

		\item
			The properties of the Wigner $d$-functions dictate that
			\begin{equation}
				\abs{\a{j,\pm{1}}}=\abs{\a{\frac{1}{2},\pm{1}}}^{2j}.
			\end{equation}	
	\end{itemize}	

	The frequencies of classical trajectories in all available $\mu$ subspaces after the quench are given by the contour integral
	\begin{equation}
		T^{(\mu)}=\oint\frac{\d q}{\dot{q}^{(\mu)}},
	\end{equation}
	where
	\begin{equation}
		\dot{q}^{(\mu)}\equiv\partialderivative{\h{\mu}}{p}=\left.p\left[1+\lambda^{2}\delta^{2}\mu\frac{1}{\sqrt{2\lambda^{2}(q^{2}+\delta^{2}p^{2})+1}}\right]\right|_{\h{\mu}(p,q,\lf)=\h{\mu}(0,q_{\mathrm{min}},\lf)}
	\end{equation}
	is the velocity of the oscillator coordinate in subspace $\mu$.
	Since the energy $e$ is measured in units $\epsilon\equiv 2jR\omega$, see Eq.~\eqref{eq:hoperator}, the period is expressed in units $\tau=\hbar/\epsilon$, where we fix $\hbar=1$ in this study.

	The frequencies are shown in the bottom row of Fig.~\ref{fig:Amplitudes}(a) for $\lambda_{i}=3/2$ and two values of $\delta$.
	Note that the critical quench $\lf^{(c)}=\sqrt{209}/12\approx1.205$ with the divergent period is not shown in the Figure.

	\begin{figure}[htbp!]
		\centering
		\includegraphics[width=\textwidth]{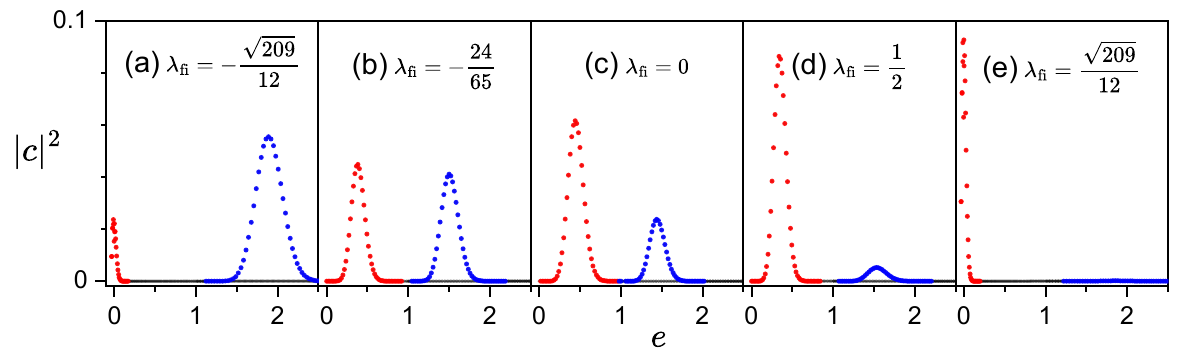}
		\caption{Components of the strength function~\eqref{eq:SF} for the Rabi model with $j=1/2$, $\li=1.5$, $R=50$ and five different quenches.
		(a,e) Critical quenches~\eqref{eq:CriticalQuench} after which the stationary point in subspace $\mu=-1$ is hit (the frequency of the classical trajectory in this subspace is infinite). 
		(b) Special quench~\eqref{eq:SameMinMaxContribution} after which both the peaks of the strength function have the same magnitude.
		(c) Quench to the noninteracting configuration.
		(d) A generic quench corresponding to the animation named \emph{Schnellbruder}~\cite{YouTubeBruder}.
		}
		\label{fig:SF1}
	\end{figure}

	\begin{figure}[htbp!]
		\centering
		\includegraphics[width=\textwidth]{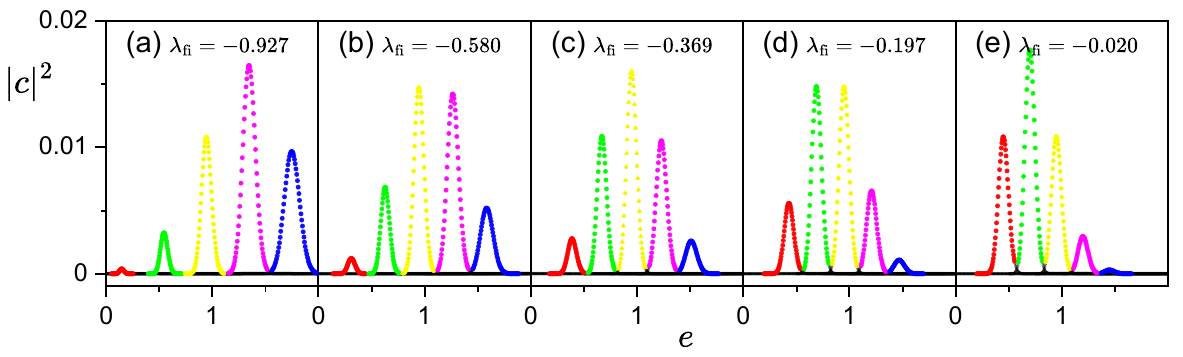}
		\caption{Components of the strength function~\eqref{eq:SF} for the Rabi model with $j=2$, $\li=1.5$ and $R=50$ and five different quenches indicated by dash-dot lines in Fig.~\ref{fig:Amplitudes}(c).
		Panel (c) corresponds to the animation of the Wigner quasiprobability distribution named 
		\emph{Five quantum brothers}~\cite{YouTubeBruder}.
		}
		\label{fig:SF4}
	\end{figure}

	Finally, let us focus on the structure of the after-quench strength function
	\begin{equation}
		\label{eq:SF}
		S(e)=\sum_{k}\underbrace{\abss{\braket{e_{k}(\lf)}{\Psi(0)}}}_{\abss{c_{k}}}\delta(e-e_{k}(\lf))
	\end{equation}
	where $\ket{e_{k}(\lf)}$ are the eigenstates of the final quantum Hamiltonian $\operator{h}(\lf)$ with corresponding rescaled energies $e_{k}(\lf)$.
	In case $j=1/2$ and $j=2$, the strength function has 2 and 5 distinguishable peaks, see Figs.~\ref{fig:SF1} and~\ref{fig:SF4}, respectively.
	The total strength of each peak, calculated as a partial sum of $\abss{c_{k}}$ contributions displayed by different colors, corresponds to the squared amplitudes $\abs{\a{\mu}}^{2}$, as also shown in the top panel of Fig.~\ref{fig:Amplitudes}.

	\section{Small coupling approximation}
	For small values of the coupling constant $\lambda\ll1$, the classical Hamiltonian~\eqref{eq:hcl} can be approximated by
	\begin{align}
		\h{\mu}(p,q)&=\frac{1}{2}\left(p^{2}+q^{2}\right)+\frac{\mu}{2}\sqrt{2\lambda^{2}\left(q^{2}+\delta^{2}p^{2}\right)+1}\nonumber\\
		&\approx\frac{1}{2}\left[p^{2}\left(1+\mu\lambda^{2}\delta^{2}\right)+q^{2}\left(1+\mu\lambda^{2}\right)\right],
	\end{align}
	which is simply a harmonic oscillator with frequency
	\begin{align}
		\omega^{(\mu)}&=\sqrt{\left(1+\mu\lambda^{2}\delta^{2}\right)\left(1+\mu\lambda^{2}\right)}\nonumber\\
		\label{eq:Weak}
		&\approx 1+\frac{\mu\lambda^{2}}{2}\left(1+\delta^{2}\right).
	\end{align}

	For $\lambda=\pm2/5=\pm0.4$ and $\delta=1/2$, the frequencies are
	\begin{equation}
		\omega^{(\mu)}\approx 1+\frac{\mu}{10},
	\end{equation}
	which gives the ratios of frequencies in all available $\mu$ subspaces $9:11$, $9:10:11$, and $18:19:20:21:22$ for $j=1/2,1$, and 2, respectively.
	The middle trajectory with $\mu=0$ (not present when $j$ is half-integer) has period $T_{0}=2\pi\approx 6.28$ (in units of $\tau$).
	The trajectories with positive $\mu$ are faster, the ones with negative $\mu$ are slower.

	In the main text of the paper, we select $\li=1.5$, $\lf=-\sqrt{2}/5\approx-0.283$, $\delta=0.5$, which gives $\omega^{(\pm1)}\approx1\pm0.05$ and periods $T^{(-1)}\approx6.60$ and $T^{(+1)}\approx5.97$.
	Therefore, the two trajectories of subspaces $\mu=-1$ and $\mu=+1$ approximately meet at their initial positions at $t\approx66$ (approximate revival) after 10 and 11 revolutions, respectively. 
	This revival is captured in the Supplemental animations~\cite{SupplementBruder}.

	\section{Approximate factorization and fast oscillations}
	After quenches to $\abs{\lf}<1$, the state of the system approximately factorizes at a specific time $\tau$,
	\begin{equation}
		\ket{\Psi(\tau)}\approx\ket{p(\tau),q(\tau)}\otimes\sum_{\mu}\alpha^{(\mu)}(\tau)\ket{\mu(\tau)},
	\end{equation}
	where $\ket{\mu(\tau)}$ are eigenstates of the third component of the quasispin operator corresponding to quantum number $m=j\mu$ in the frame where the effective magnetic field $\Bf(\tau)$ points along the $z$-axis.
	If we assume that $R\gg1$, so that the spin motion is much faster and the oscillator can be adiabatically separated, then Hamiltonian~\eqref{eq:h} induces evolution
	\begin{equation}
		\ket{\Psi(\tau+t)}=\ket{p(\tau),q(\tau)}\otimes\sum_{\mu}\a{\mu}(\tau)\e^{\im\mu\abs{\Bf(\tau)}t}\ket{\mu(\tau)}.
	\end{equation}
	For a particular choice $j=1/2$,
	\begin{equation}
		\ket{\Psi(\tau+t)}=\ket{p(\tau),q(\tau)}\otimes\left[\a{-1}(\tau)\e^{\im\abs{\Bf(\tau)}t}\ket{-1(\tau)}+\a{+1}(\tau)\e^{-\im\abs{\Bf(\tau)}t}\ket{+1(\tau)}\right],
	\end{equation}
	so the ``local survival probability'' 
	\begin{equation}
		P_{\tau}(t)\equiv\abss{\braket{\Psi(\tau)}{\Psi(\tau+t)}}=1-4\abss{\a{-1}(\tau)\a{+1}(\tau)}\sin^{2}\left(\abs{\Bf(\tau)}t\right)
	\end{equation}
	oscillates with the Rabi frequency 
	\begin{equation}
		\omega_\tau=\abs{\Bf(\tau)}\approx\frac{1}{2}\left\{1+\lf^{2}\left[q^{2}(\tau)+\delta^{2}p^{2}(\tau)\right]\right\}
	\end{equation}
	With the same frequency oscillates also the expectation values of the quasispin operator components~\cite{YouTubeBruder,SupplementBruder}
	\begin{equation}
		j_{k,\tau}(t)\equiv\matrixelement{\Psi(\tau+t)}{\operator{j}_{k}}{\Psi(\tau+t)}.
	\end{equation}

	\section{Purity and Wigner negativity}
	The purity of the spin-oscillator system is defined as
	\begin{equation}
		\gamma(t)=\trace{\operator{\rho}_{\mathrm{osc}}^{2}(t)},
	\end{equation}
	where $\operator{\rho}_{\mathrm{osc}}(t)=\trace_{\mathrm{sp}}{\operator{\rho}}(t)$ is the reduced density matrix of the oscillator obtained by tracing out the spin degrees of freedom from the total density matrix $\operator{\rho}(t)=\ket{\Psi(t)}\bra{\Psi(t)}$.
	It ranges from $\gamma=1$ for a pure state to $\gamma=1/(2j+1)$ for a maximally mixed state.

	The Wigner negativity is given by
	\begin{equation}
		\nu(t)=\frac{1}{2}\left[\int\abs{W_{\mathrm{osc}}(q,p,t)}dq\,dp-1\right],
	\end{equation}
	where $W_{\mathrm{osc}}(q,p,t)$ is the Wigner quasiprobability distribution of the oscillator at time $t$.

	\begin{figure}[htbp!]
		\centering
		\includegraphics[width=\textwidth]{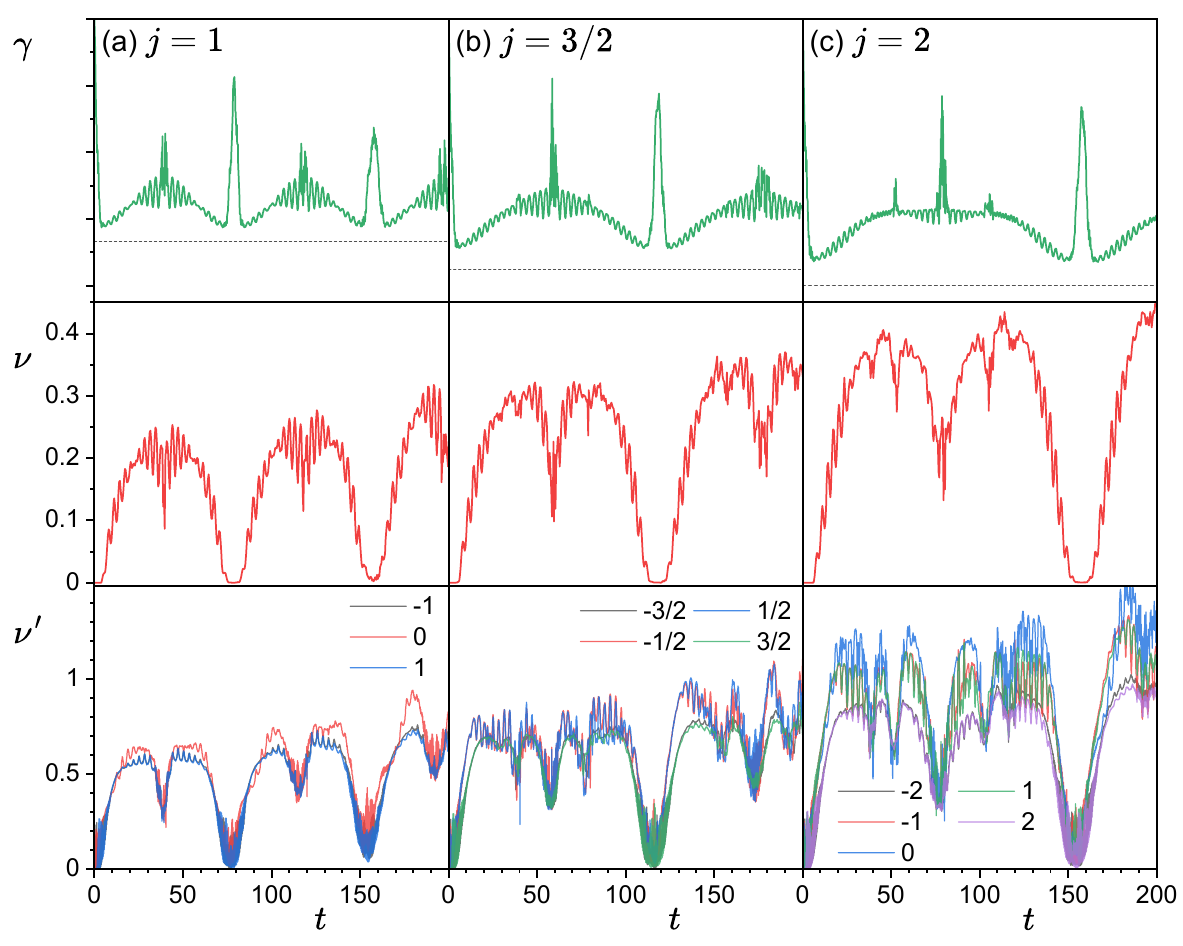}
		\caption{Purity (first row) and negativity (second row) after quench $\li=1.5\rightarrow\lf=-0.369$ of the Rabi model with $R=50$ and various spin sizes. 
		Horizontal dashed lines in the purity panels indicate the minimum purity $1/(2j+1)$ for a maximally mixed state.
		Third row: Wigner negativity immediately after performing a measurement of the spin state in direction I defined in Sec.~\ref{sec:Measurement}.
		Different colors correspond to different measurement spin projections.
		}
		\label{fig:PNj}
	\end{figure}	

	Both the purity and negativity are displayed in the first two rows of Fig.~\ref{fig:PNj} for quench $\li=1.5\rightarrow\lf=-0.369$ and various values of~$j$.
	We can see that the time evolution of the purity has a general trend: after an initial decay, it rises again and oscillates around a value that depends on $j$ only weakly, hence increases relatively to the minimum purity $1/(2j+1)$ with increasing $j$, at least for small values of $j$.
	When the system is close to a revival, the purity reaches values close to 1, indicating that the spin and oscillator are nearly disentangled at these times.
	Due to gradual spreading of the wave packet, the revivals become less and less perfect for longer times.

	On the other hand, the negativity vanishes at $t=0$ and at the revivals, and is positive in the windows between the revivals. 
	Its mean value in the inter-revival windows increases with $j$, which is caused by the increasing number $2j+1$ of the wavepackets (more components of the cat state) because each pair of subspace wavepackets creates one region of interference fringes with pronounced negativities and the number of wavepackets pairs grows quadratically with $2j+1$.
	The negativity also increases with time, which is caused by the wavepacket spreading and approaching state equilibration with necessary quantum fluctuations.

	\begin{figure}[htbp!]
		\centering
		\includegraphics[width=\textwidth]{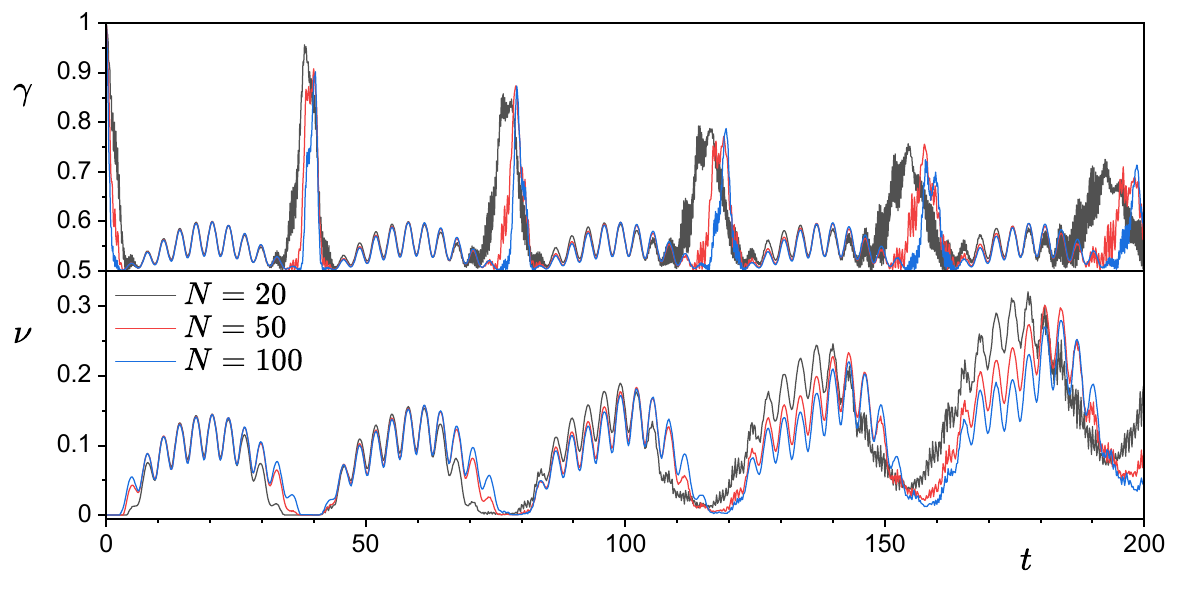}
		\caption{Scaling of purity and negativity with size parameter $R$ for quench $\li=1.5\rightarrow\lf=-0.369$ of the Rabi model with $j=1/2$.
		}
		\label{fig:Scaling}
	\end{figure}	

	Next, let us discuss the scaling of the purity and negativity with the size parameter $R$, shown in Fig.~\ref{fig:Scaling}.
	We observe that at the beginning of the evolution, the purity decays faster and the negativity starts rising sooner for larger $R$. 
	However, between the initial decay and the first revival and in the subsequent inter-revival windows, both quantities seem to equilibrate around values that are almost independent of $R$, at least before the wavepacket spreading becomes significant.
	Note that the negativity is affected more strongly by the spreading due to increasing importance of the quantum fluctuations.
	The higher $R$, the narrower the wave packet and the later the appearance of the spreading effects.

	\begin{figure}[htbp!]
		\centering
		\includegraphics[width=\textwidth]{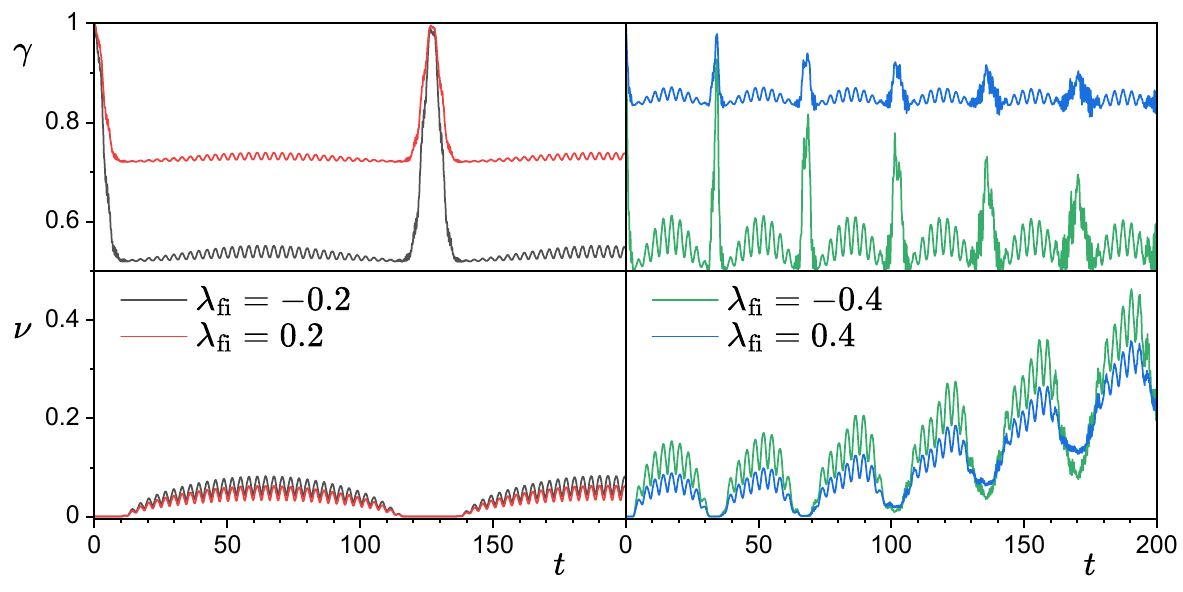}
		\caption{Purity and negativity for various quenches $\li=1.5\rightarrow\lf$ of the Rabi model with $j=1/2$ and $R=50$.
		}
		\label{fig:Lambda}
	\end{figure}	

	Finally, we present the purity and negativity for various values of the after-quench coupling $\lf$ in Fig.~\ref{fig:Lambda}.
	It is observed that the revival times, indicated by $\gamma\approx 1$ and $\nu\approx 0$, and frequencies of the inter-revival oscillations are independent of the sign of $\lf$---a fact related to the symmetry $\h{\mu}(\lambda)=\h{\mu}(-\lambda)$ of the semiclassical Hamiltonian~\eqref{eq:hcl}.
	However, the weights of the individual subspace contributions in the after-quench state depend on $\lf$, recall Eq.~\eqref{eq:alpha12}, Fig.~\ref{fig:Amplitudes} and the strength function in Fig.~\ref{fig:SF1}.
	This leads to higher average the purity and negativity inside the inter-revival time intervals for quenches with more balanced contributions from both subspaces, \ie quenches with $\lf$ close to $\lf^{(=)}$, because it is the interference between the subspace wavepackets that makes the state nonclassical with higher values of the Wigner negativity.
	The figure also shows that the quenches to very small values of $\abs{\lf}$ lead to very long revival times, in accordance with the small-coupling approximation discussed above, and at the same time to slower wavepacket spreading and equilibration.

	\section{Measurement as state purification}
	\label{sec:Measurement}
	As seen in the previous section, the purity of the oscillator subsystem is generally less than 1 at times between the revivals when the multicomponent structure of the state is well-developed.
	However, the state can be purified by performing a projective measurement of the spin state.
	We employ the following protocol: after the quench, we evolve state~\eqref{eq:alphagen} into an entangled state at time $t$,
	\begin{equation}
		\ket{\Psi(t)}\approx\sum_{m=-j}^{j}\alpha^{(j,m)}\ket{\psi^{(m)}(t)}\otimes\ket{m(t)},
	\end{equation}
	where $\ket{m(t)}\equiv\ket{m;\vector{n}^{(m)}(t)}$ and the directions $\vector{n}^{(m)}(t)$ are determined from the effective magnetic field~\eqref{eq:B} by inserting averages
	\begin{equation}
		q^{(m)}(t)=\matrixelement{\psi^{(m)}(t)}{\operator{q}}{\psi^{(m)}(t)},\quad p^{(m)}(t)=\matrixelement{\psi^{(m)}(t)}{\operator{p}}{\psi^{(m)}(t)}.
	\end{equation}
	Note that the states $\ket{m(t)}$ are not mutually orthogonal; for a more detailed discussion see the main text.
	After that, we measure the spin projection in a direction described by angles $\theta',\phi'$ in the laboratory coordinate frame.
	The measurement outcome is $m'=-j,-j+1,\dotsc,j$ and the post-measurement factorized state is
	\begin{equation}
		\ket{\Psi'_{m'}(t)}=\mathcal{N}\left[\sum_{m=-j}^{j}\alpha^{(j,m)}\braket{m'}{m(t)}\ket{\psi^{(m)}(t)}\right]\otimes\ket{m'},
	\end{equation}
	where $\mathcal{N}$ is a normalization constant and $\ket{m'}\equiv\ket{m';\vector{n}'}$ is the eigenstate of the spin operator in the measurement direction $\vector{n}'=(\sin\theta'\cos\phi',\sin\theta'\sin\phi',\cos\theta')$ corresponding to eigenvalue $m'$.
	The scalar product $\braket{m'}{m(t)}$ can, in principle, be calculated from the Wigner $d$-functions.

	\begin{figure}[htbp!]
		\centering
		\includegraphics[width=\textwidth]{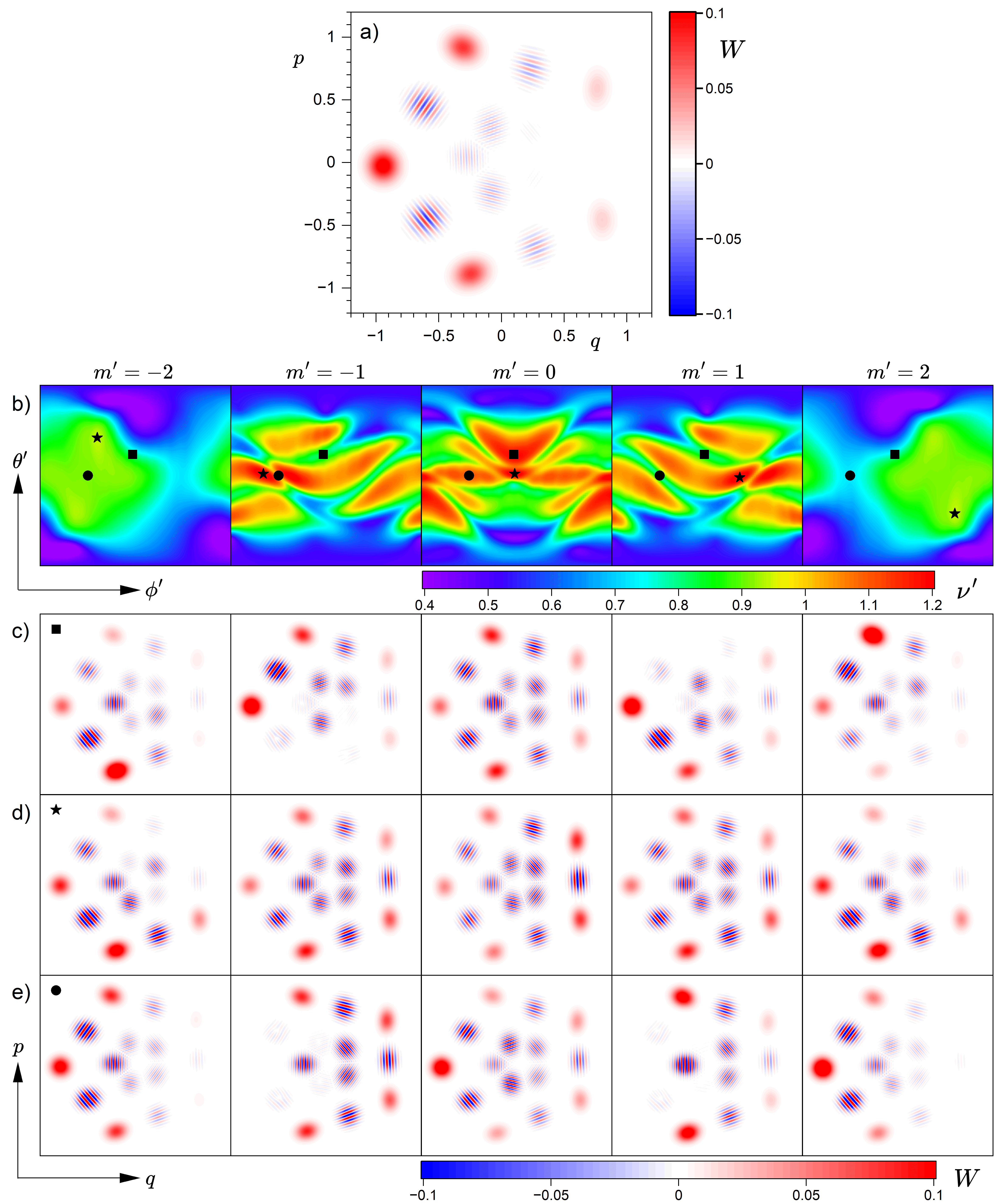}

		\caption{(a) Entangled state $\ket{\Psi(t)}$ at time $t=31.4$ for the Rabi model with $j=2$, $\li=1.5$, $\lf=-0.369$ and $R=50$.
		(b) Wigner negativity following a measurement in the direction $(\theta',\phi')$, shown for all possible measurement outcomes $m'=-2,-1,0,1,2$.
		The square, star and bullet symbols mark measurement directions I, II and III, respectively.
		The Wigner quasiprobability distributions of the resulting post-measurement states are displayed in panels (c), (d), and (e), respectively.
		}
		\label{fig:Measurement}
	\end{figure}	

	An example of the oscillator Wigner quasiprobability distribution before and after performing the measurement in different directions is shown in Fig.~\ref{fig:Measurement} for $j=2$ and the same quench as in the previous sections.	
	Panel (a) shows the entangled state $\ket{\Psi(t)}$ at time $t=31.4$ when the multicomponent structure is well-developed. However, the purity is relatively low, see Fig.~\ref{fig:PNj}.
	Therefore, we perform the measurement of the spin to purify the state and enhance its nonclassical features.
	The Wigner negativity of the post-measurement state is shown in row (b) for all possible measurement outcomes $m'=-2,-1,0,1,2$ and for all possible measurement directions $(\theta',\phi')$.
	The Wigner quasiprobability distributions of the post-measurement states for three particular directions are shown in the last three rows of Fig.~\ref{fig:Measurement}: Row (c) chooses the direction along the vector $\vector{n}^{(m=-2)}(t)\times\vector{n}^{(m=2)}(t)$ (direction I, square symbol), which is perpendicular to the plane spanned by the two most distant spin states $\ket{m(t)=-2}$ and $\ket{m(t)=2}$; this is equivalent to the direction employed in the main text. 
	Row (d) chooses the direction that maximizes the Wigner negativity for the corresponding outcome $m'$ (direction II, star symbol): note that for each outcome $m'$, this direction is different.
	Finally, row (e) chooses a fixed measurement direction along the $y$-axis in the laboratory frame (direction III, bullet symbol).

	We observe that the measurement procedure does not affect the overall structure of the Wigner quasiprobability distribution.
	It slightly changes the relative weights of the contributions from different $m$ subspaces, and it enhances the interference fringes between them, which leads to an increase of the Wigner negativity, as shown in the last row of Fig.~\ref{fig:PNj}.
	However, the measurement direction does not significantly affect the Wigner negativity and the Wigner quasiprobability distribution of the post-measurement state, as shown in Fig.~\ref{fig:Measurement}.
	
	A more detailed study shows that for similar quenches to the one presented here, any direction in the $xy$ plane in the laboratory frame gives suitable and consistent nonclassical states with similar Wigner negativities in all measurement outcomes, cf. Fig.~\ref{fig:Measurement}(b).

	\section{Summary}
	We conclude this Supplemental material by summarizing the protocol for the generation of pure nonclassical states of the oscillator subsystem in the Rabi model:
	\begin{enumerate}
		\item Prepare the system with spin $j$ in the parity violating ground state of the initial Hamiltonian $\operator{h}(\li)$, $\abs{\li}>\abs{\lambda_{c}}$, which is approximately given by a product state of a displaced coherent state of the oscillator and a spin state pointing in the direction of the initial effective magnetic field $\Bi$, see Eq.~\eqref{eq:InitialState}.
		\item Perform a quench of the coupling constant $\lambda$ from $\li$ to $\lf$, which leads to an after-quench state that can be expressed as a superposition of states localized in different subspaces $\mu=m/j$.
		\item Evolve the system up to time $t$, until a multicomponent structure of the state is well-developed.
		\item Optionally, switch off the spin-oscillator interaction by setting $\lambda=0$, which freezes relative positions and shapes of the wavepackets in different subspaces $\mu$ and their mutual interference pattern.
		\item Perform a projective measurement of the spin state in a chosen direction, which purifies the oscillator state and can enhance its nonclassical features.
	\end{enumerate}

	The Rabi model has the following parameters that can be tuned to prepare various configurations of final nonclassical states:
	\begin{itemize}
		\item 
			$j$ determines the number of subspaces $\mu$ and therefore the number of components $2j+1$ in the after-quench state, see Eq.~\eqref{eq:alphagen} and Fig.~\ref{fig:Amplitudes}.
		\item 
			$\li$ and $\lf$ determine the weights of the contributions from different $\mu$ subspaces in the after-quench state, see Eq.~\eqref{eq:alphagen} and Fig.~\ref{fig:Amplitudes}, their frequencies, see Fig.~\ref{fig:Amplitudes}, and their stability.
			For small $\abs{\lf}$, the Hamiltonian can be approximated by a harmonic oscillator with slightly different frequencies in different $\mu$ subspaces, see Eq.~\eqref{eq:Weak}, which leads to long revival times and high wavepacket stability caused by their slow spreading.
		\item
			$\delta$ affects the frequencies of the trajectories in different $\mu$ subspaces, see Eq.~\eqref{eq:Weak} and Fig.~\ref{fig:Amplitudes}.
		\item
			The time $t$ of the evolution after the quench determines the degree of development of the multicomponent structure of the state and therefore its nonclassical features, as well as the degree of wavepacket spreading and equilibration.
		\item
			The measurement direction, described by angles $(\theta',\phi')$, determines the final structure of the post-measurement state, see Fig.~\ref{fig:Measurement}.
	\end{itemize}

	\section{Supplemental animations}
	There are two independent lists of animations demonstrating the dynamics of the Wigner function and mean values of some operators after the quench.
	All the animations share the following parameters: $\li=1.5$, $\delta=1/2$.

	First, the animations corresponding to the quenches presented in the main text of the paper are shown in the Supplemental material~\cite{SupplementBruder}.
	These animations have $\lf=-0.283$ and size $R=20$ and differ by the value of $j$: $j=1/2,1,2$.

	Second, additional animations are presented in a dedicated YouTube channel~\cite{YouTubeBruder}.
	These animations have been made more attractive and funny by adding AI generated music with AI written lyrics in various languages.
	They are:
	\begin{itemize}
		\item \href{https://youtu.be/xajXXezmg48}{Schnellbruder}: $j=1/2$, $\lf=0.5$.
		\item \href{https://youtu.be/guQtmg9V7p8}{Dos gotas de Rabi}: $j=1/2$, $\lf=\lf^{(c)}\approx1.205$ (the wavepacket in $\mu=-1$ subspace eventually hits the stationary point at $p=q=0$).
		\item \href{https://youtu.be/Fnqfdxmvj8w}{Rabiho rezonance dvou}: $j=1/2$, $\lf=-\li=-1.5$ (the wavepacket in $\mu=-1$ subspace remains sitting in the initial minimum at $p=0,q=q_{\mathrm{min}}$).
		\item \href{https://youtu.be/Bv5ec2nnh7E}{Three quantum brothers}: $j=1$, $\lf=\lf^{(=)}\approx0.369$.
		\item \href{https://youtu.be/Bv5ec2nnh7E}{Five quantum brothers}: $j=2$, $\lf=\lf^{(=)}\approx0.369$.
	\end{itemize}

	The layout of all the animations (both from the Supplemental material and YouTube) is the same and contains the following panels:
	\begin{itemize}
		\item The main panel shows the Wigner quasiprobability distribution of the oscillator with its $q$ and $p$ marginals (squared moduli of the wave functions in $q$ and $p$ representations) displayed on the top and right sides of the panel, respectively.
			Time $t$ is given in the panel caption.
			The bullets indicate the positions of the classical oscillator (solution of equations of motion given by the Hamiltonian~\eqref{eq:hcl}) in all available $\mu$ subspaces.
			The lines preceding the bullets show the last 10 time units (expressed in units of $\tau$) of the classical trajectories.
			The classical evolution is projected onto the Bloch sphere (right-bottom panel) using the classical version of Eq.~\eqref{eq:B}.
		\item The panels labelled $J_{x},J_{y},J_{z}$ show the quantum expectation values of the quasispin operator components,
			\begin{equation}
				J_{k}(t)=\matrixelement{\Psi(t)}{\operator{J}_{k}}{\Psi(t)},\quad k=x,y,z.
			\end{equation}
			The vector whose components are given by scaled values $J_{k}(t)/j$ is also displayed on the Bloch sphere (right-bottom panel, gray diamond and gray line).
		\item The panels labelled $q,p$ show the expectation values of the quantum oscillator coordinates and momenta,
			\begin{equation}
				q(t)=\matrixelement{\Psi(t)}{\operator{q}}{\Psi(t)},\quad p(t)=\matrixelement{\Psi(t)}{\operator{p}}{\Psi(t)}.
			\end{equation}
			The corresponding normalized vector $\hat{\vector{n}}$ determined by Eq.~\eqref{eq:B} is shown on the Bloch sphere as the red square and its history as the red line.
		\item The panel labelled $P$ shows the survival probability
			\begin{equation}
				P(t)=\abss{\matrixelement{\Psi(0)}{\e^{-\im\operator{h}(\lf)t}}{\Psi(0)}}.
			\end{equation}
	\end{itemize}

\bibliographystyle{apsrev4-2}  
\bibliography{SM.bib}      